# Biological control networks suggest the use of biomimetic sets for combinatorial therapies

Jacob D. Feala[1], Jorge Cortes[2], Phillip M. Duxbury[3], Andrew D. McCulloch[4], Carlo Piermarocchi[3], Giovanni Paternostro[1]

[1]Sanford-Burnham Medical Research Institute, La Jolla, California
[2]Department of Mechanical and Aerospace Engineering, University of California, San Diego
[3]Department of Physics and Astronomy, Michigan State University
[4]Department of Bioengineering, University of California, San Diego

Please address correspondence to
Giovanni Paternostro
e-mail: giovanni@burnham.org

and to

Jacob Feala
e-mail: jfeala@gmail.com

Further discussion can be found at http://www.combiocontrol.org

**Summary:** The design of drug combinations may benefit from a novel, network-level approach that mimics control structures found across a variety of biological systems.

**Abstract**

Cells are regulated by networks of *controllers* having many targets, and *targets* affected by many controllers, but these "many-to-many" combinatorial control systems are poorly understood. Here we analyze distinct cellular networks (transcription factors, microRNAs, and protein kinases) and a drug-target network. Certain network properties seem universal across systems and species, suggesting the existence of common control strategies in biology. The number of controllers is ~8% of targets and the density of links is 2.5% ± 1.2%. Links per node are predominantly exponentially distributed, implying conservation of the average, which we explain using a mathematical model of robustness in control networks. These findings suggest that optimal pharmacological strategies may benefit from a similar, *many-to-many* combinatorial structure, and molecular tools are available to test this approach.

Control of cellular function depends on bipartite networks, in which one class of nodes (the controller) acts on the other class (the target) to regulate its function. Examples of cellular control networks include transcription factors, microRNAs, and protein kinases. In these networks, the control layer interacts with the target layer in a combinatorial, "many-to-many" fashion (see Figure 1). In other words, each controller has many targets, the targets themselves are under the influence of many controlling molecules, and the target sets of different controllers overlap. Moreover, the number of controllers is usually significantly lower than the number of targets. While this is well recognized in biological systems (see supplementary online material), cellular combinatorial control is not well understood at a systems level. This limits our ability to design more natural modes of biological intervention. Here we attempt to quantify general features of combinatorial control networks, with the aim of discovering properties that can be used to develop a new, biomimetic paradigm for pharmacological control of disease.

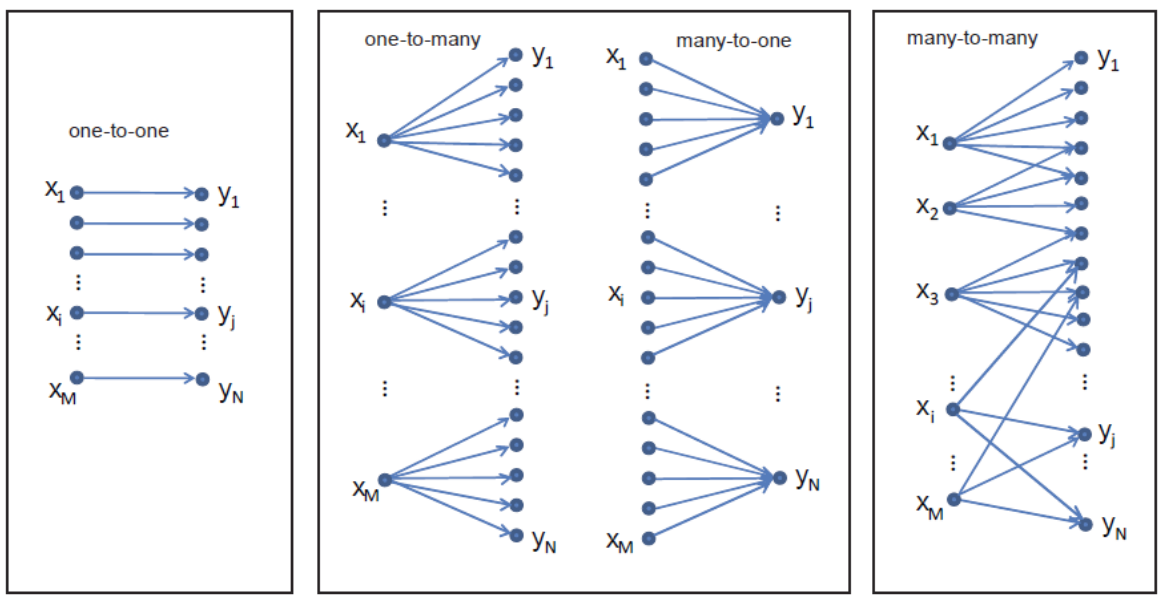

**Figure 1: Possible combinatorial control strategies.** There are several qualitatively different structures for control networks of M controllers ($x_1, x_2, ... x_M$) and N targets ($y_1, y_2, ... y_N$). In the one-to-one case (left panel), M = N.

We examine quantitative characteristics of these three biological control systems in three different species (human, yeast, and *E. coli*), from the perspective of two-layer combinatorial control. First we consider the numbers of nodes. Table 1 (top left) shows estimates of the number of controllers and targets from the literature for the three types of networks in humans. Notably, though these numbers are from three different cellular systems of varying size, the ratios of control nodes to target nodes are similar, approximately 8% (Table 1, upper left).

Next, we use molecular interaction databases to explore connectivity parameters of bipartite networks in nature. Networks were extracted from publicly available databases and separated into controller nodes (microRNA, transcription factors, protein kinases) and target nodes (mRNA transcript, gene, phosphorylated protein substrate), with directed links between controllers and targets. We also used existing protein kinase target profiles for a set of kinase inhibitors (*1*) (KIs), for comparison with the endogenous networks. While there have been many genome-wide network analyses (*2-7*), and one recent work on co-regulation of transcription and phosphorylation networks (*8*), here we focus on universal features of bipartite networks that may help design biomimetic control strategies. We quantified properties including density of links (existing links divided by the number of possible links), distribution of links for each type of node, and overlap between the target sets of different controllers. Where possible, we gauged the biological significance of these features by comparison with those expected from a random network having the same number of nodes and links. Deviations from random may indicate a functional advantage for a particular network characteristic.

Table 1 shows that these networks share specific network-wide properties despite wide variation in the number of nodes, complexity of species, and type of molecular interaction. The mean M/N over all biological networks was 8.9%. Detailed analysis of the measures of overlap (Shared Targets per Controller and Pairwise Overlap) and Gene Ontology (GO) enrichment of targets are in the supplementary material.

Figure 2 shows distributions of links per node $k$, for incoming links (controllers per target, $k_{in}$) and outgoing links (targets per controller, $k_{out}$). Figure 2A depicts the empirical cumulative distribution function (cdf) for all datasets, normalized by the average links per node *<k>* and overlaid on a standard exponential cdf (solid line). Figures 2B and 2C show histograms of each individual network, compared with the binomial distributions expected of random networks of the same size generated by the Erdös-Rényi random graph model (*9*). The human transcription factor network has a peak in its outgoing link distribution that is accurately approximated by a binomial distribution. The incoming links in the kinase inhibitor network also show a possible binomial component. Otherwise, most curves closely approximate an exponential distribution (more details in the Supplementary Online Material).

**Table 1: Network parameters for various types of combinatorial control within cells.** The ratio of controllers per target drawn from the literature is similar across different types of biological network in humans, approximately 8%. Node properties differ between the literature and network databases owing to incomplete information in the databases. Link density is the ratio of the number of actual links to the number of possible links. Shared targets per controller and pairwise overlap are measurements of overlapping target sets described in the Supplementary Online Material. SD = standard deviation, CV = coefficient of variation.

|  | Literature | | | Network databases | | | | | | |
| --- | --- | --- | --- | --- | --- | --- | --- | --- | --- | --- |
|  | Human | | | Human | | | Yeast | | E. coli | Drug |
| **Node properties** | TF | Kinase | miRNA | TF | Kinase | miRNA | TF | Kinase | TF | KI |
| Controllers (M) | 1,800* | 518 | 940 | 389 | 264 | 153 | 186 | 88 | 169 | 38 |
| Targets (N) | 20,500• | 6,150‡ | 11,890† | 9284 | 988 | 9448 | 6297 | 1341 | 1495 | 316 |
| M / N (%) | 8.8% | 8.4% | 7.9% | 4.2% | 26.7% | 1.6% | 3.0% | 6.6% | 11.3% | 12.0% |
| **Link properties** | | | | | | | | | | |
| Targets per controller (mean) | | | | 181 | 8.9 | 359 | 229 | 46 | 20 | 78.8 |
| Controllers per target (mean) | | | | 7.6 | 2.4 | 5.8 | 6.8 | 3 | 2.3 | 9.48 |
| Link density | | | | 1.9% | 0.9% | 3.5% | 3.6% | 3.5% | 1.3% | 25.0% |
| Shared targets per controller (mean) | | | | 98% | 73% | 95% | 98% | 85% | 74% | 100% |
| Pairwise overlap of targets (mean) | | | | 4.5% | 7.1% | 7.1% | 6.3% | 8.3% | 1.1% | 33.8% |

**Statistical values of selected parameters**

| Parameter | Mean | SD | CV | 95% lo | 95% hi |
| --- | --- | --- | --- | --- | --- |
| M/N (literature) | 8.4% | 0.5% | 0.054 | 7.5% | 9.3% |
| Controllers per target | 4.7 | 2.4 | 0.51 | 0.016 | 9.3 |
| Link density | 2.5% | 1.2% | 0.50 | 0.041% | 4.9% |
| Shared targets per controller | 87.2% | 11.6% | 0.13 | | |
| Pairwise overlap of targets | 5.7% | 2.6% | 0.45 | 0.6% | 10.8% |

*Vaqueriza et. al. estimate 1,700-1,800 human transcription factors
•Other estimates for the number of human genes are in the range 20,000 - 25,000
†Friedman et. al. estimate 58% of genes are targeted by miRNA (11,890 = .58*20,500)
‡Cohen et. al. estimate 30% of human proteins are phosphoryated (6,150 = .30*20,500)

The "maximum entropy distribution," for a random variable with specified mean and unspecified variance, is an exponential distribution (*10*). For example, the height of a molecule in a gas under gravitational force has its mean constrained but no limit to its variance, and thus is exponentially distributed under thermodynamic equilibrium.

Therefore, any evolving system that conserves the mean but not the variance of some observable sequence of variables (i.e., links per node), will maximize entropy of that sequence over time, regardless of the underlying processes, and eventually reach an exponential distribution (*11*). This is simply because these distributions are vastly more common in the space of all possible sequences with specified mean and unspecified variance. The observed distributions imply that the average links per node is tightly conserved (specified mean), while the number of links in any single node is unlimited (unspecified variance).

Other genome-wide interaction networks, such as protein-protein interaction, metabolic, and gene co-expression networks, also have distributions of links that deviate from the random graph case (*2-5*), but these are generally scale-free rather than exponential distributions. Scale-free distributions can also arise by maximum entropy with unconstrained variance, but these require conservation of the geometric, not the arithmetic, mean (11). We ensured that exponential distributions cannot be obtained simply by sampling a larger, scale-free network (supplementary online material).

All biological networks had similar sparse link density, realizing an average of only 2.5% ± 1.2% of all possible controller-to-target interactions. Link density $D$ is related to the average links per node by the equation

$$(1) \quad D = \frac{\langle k_{in} \rangle}{M} = \frac{\langle k_{out} \rangle}{N},$$

where $\langle k_{in} \rangle$ is the average incoming links over $N$ target nodes, and $\langle k_{out} \rangle$ is the average outgoing links from $M$ controller nodes. Note that

$$(2) \quad \frac{\langle k_{in} \rangle}{\langle k_{out} \rangle} = \frac{M}{N},$$

suggesting that similarities in the ratios of nodes may be related to constraints on the average incoming and outgoing links per node.

**Figure 2 (next page): Distributions of incoming and outgoing links** for several types of combinatorial control networks. (A) Cumulative distributions of links per node in each of the networks of Table 1 were normalized by the mean and plotted together on log-log axes, alongside the discrete analog to the exponential distribution (solid line), see Methods. By contrast, a power-law, or scale-free, distribution would produce a straight line in this log-log plot. (B) Individual histograms of targets per controller and (C) controllers per target plotted for each individual network. The three human networks were combined based on shared targets (top right of each panel). Horizontal axes in (B) and (C) are normalized to the total number of target or controller nodes, respectively in each network. Each distribution is compared with the binomial distribution expected from a random graph with identical numbers of nodes and links (dashed curve). An exponential curve is also fitted to each dataset (solid line). Note that the kinase inhibitor network is distributed over a much wider range on the x-axis than the biological networks.

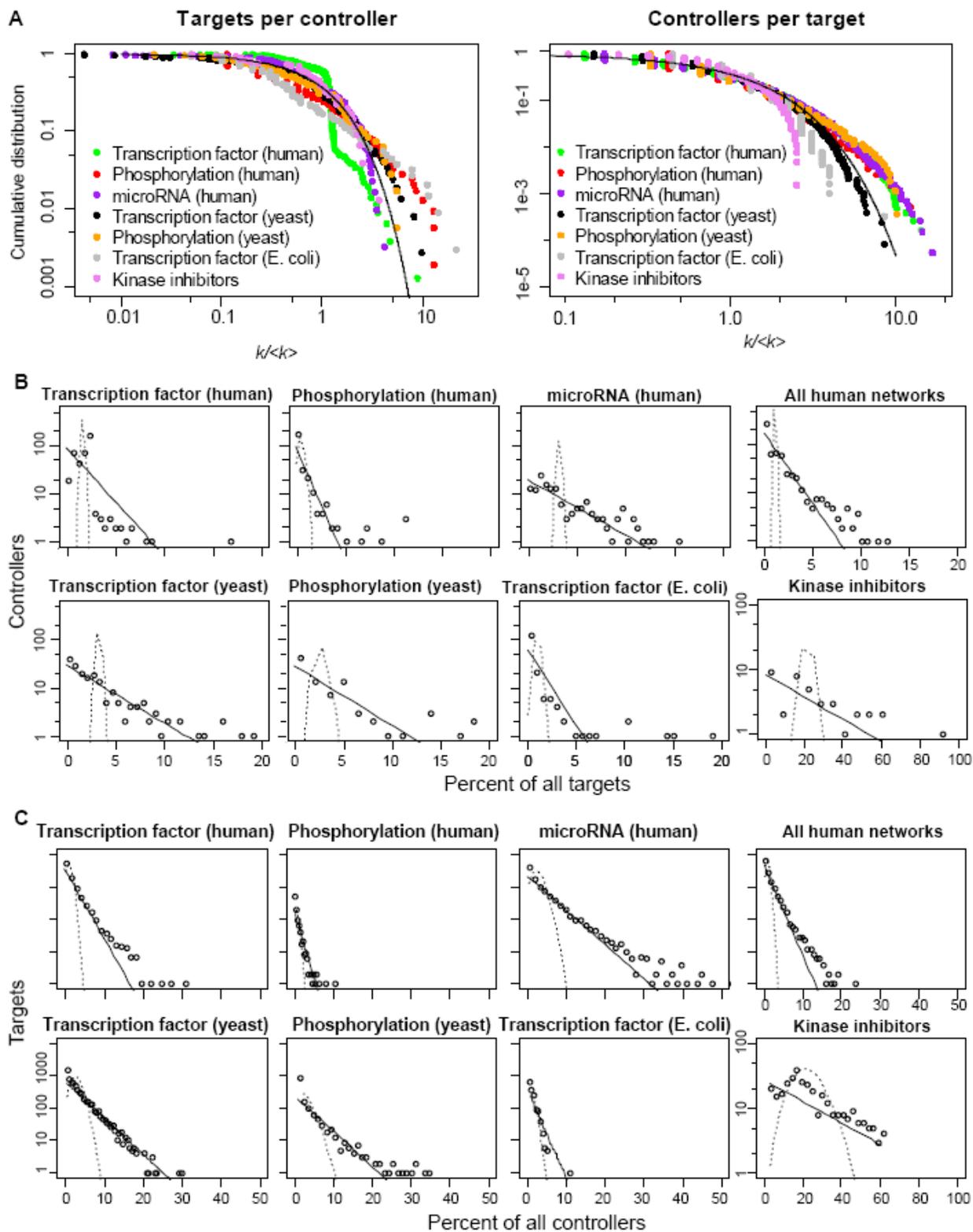

This many-to-many network structure, with parameters spanning comparatively limited ranges, may be the result of an optimized trade-off between efficient use of biological resources and robustness to variation in environmental and genetic inputs (via redundancy). To maximize redundancy, a high average incoming link degree is clearly preferable. We built an abstract model to simulate redundancy and robustness in a bipartite information processing network. A set of transcription factors, for example, takes on its expression state according to upstream signaling events, and induces an output gene expression state through its network of targets. Now consider a set of $M$ controller nodes, which can take on $2^M$ binary states. Controllers are randomly connected to $N$ target nodes having average incoming links $<k_{in}>$, and each target node takes on a binary state according to a Boolean rule (see Methods). We can then derive the number of unique output sequences $\Omega$ that the network can achieve, and the robustness $R$ of an output state to mutations (link deletions), given values of $M$, $N$, and $<k_{in}>$.

In Figure 3, analytical solutions for $\Omega$ and $R$ are plotted as a function of $<k_{in}>$ over the biological ranges of Table 1, alongside numerical simulations (see Methods). Results are similar regardless of the node computation rule or link distribution (see supplementary online material). The number of unique output states $\Omega$ is a decreasing function of $<k_{in}>$, and robustness $R$ is an increasing function of $<k_{in}>$ dependent on the mutation rate. Furthermore, $R$ increases rapidly with $<k_{in}>$ above 1, but saturates quickly for values above 5. Therefore, adding redundancy via $<k_{in}>$ has a high marginal benefit to robustness for low $<k_{in}>$, but as $<k_{in}>$ increases, the incremental benefit to $R$ may be outweighed by the cost to the unique outputs achievable by the network. There may be an additional evolutionary cost for attaining and storing the genetic information required for each link, and increasing the numbers of controllers and links may also incur a cellular cost for resources dedicated to protein synthesis. Many-to-many configurations would therefore be expected to emerge as a strategy for maximizing both robustness and the efficient use of resources, and observed network parameters reflect a balance between these opposing influences. These considerations are consistent with the differences in values of $<k_{in}>$ among human and bacterial transcription factor networks (Table 1). As pointed out by r/K selection theory (*12*), these two organisms use very different life history strategies, with bacteria favoring more rapid reproduction (facilitated by a smaller genome size) and lower offspring robustness.

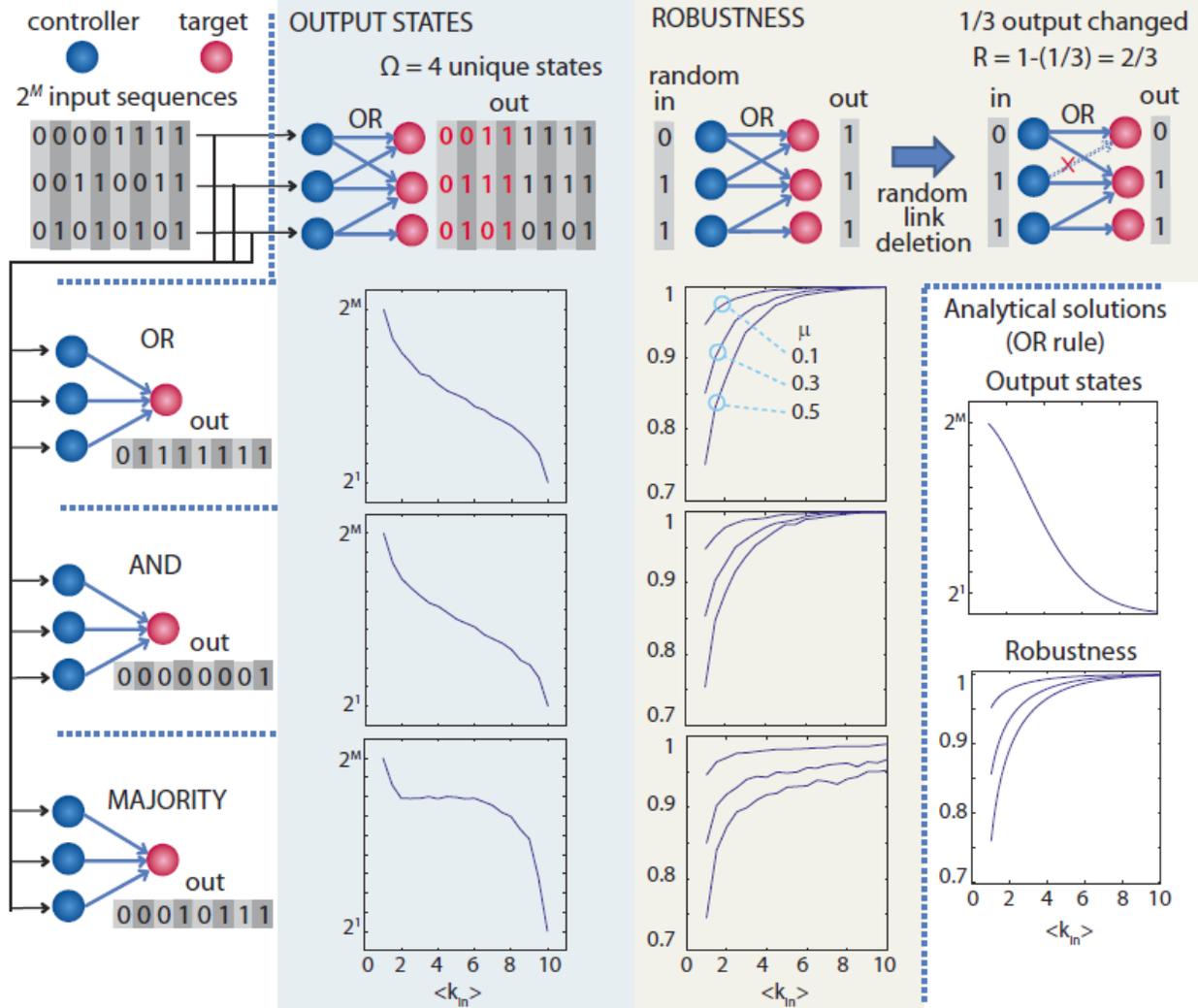

**Figure 3: Mathematical model of the number and robustness of output states** in a bipartite control network. We explored the dependence of these quantities on the average incoming links <$k_{in}$>, number of controllers $M$, number of targets $N$, and mutation rate $\mu$ (or links deleted as a fraction of $N$, robustness equation only). Shown are averages of 1000 numerical simulations with $M = N = 10$, and $\mu = 0.1$. Analytical solutions for robustness and unique output states using the OR rule were also derived and plotted (lower right), and found to be identical or a close approximation to simulations, respectively (see Methods). Both quantities were independent of $N$ in numerical and analytical solutions. These results suggest that marginal utility to robustness of increasing <$k_{in}$> shrinks rapidly above ~5, while at the same time incurring a cost on the degree of freedom of output states.

One limitation of this analysis is that the bipartite model is only a first approximation of reality, since many nodes in the target layer are controllers themselves, interactions downstream of the targets can feed back to the control layer, and nodes often do interact with other nodes of the same class. It should also be noted that these interaction datasets are incomplete, have varying levels of confidence, and are not fully validated. The quantitative patterns we have described are, however, common to datasets of very different origin and therefore cannot reasonably be explained by experimental noise or bias present in each dataset.

The evolutionarily conserved nature of the many-to-many structure (see also Supplementary Online Material for other examples in biology), certain measured network parameters, and the exponential distribution of links suggests that pharmacological control strategies should be designed similarly. Current efforts to develop specific, targeted therapies follow the one-to-one approach to drug therapy (*13, 14*); in other words, the ideal aim of drug discovery is seen as having one drug for each molecular target, with no target overlap. More traditional therapies are often less specific (one-to-many in Figure 1) and some effective targeted therapies have also been found to be non-specific and might fit this category (*15, 16*).

An alternative approach would seek combinations of drugs that control the activation state of a set of targets in a many-to-many fashion, similar to combinatorial regulation of cellular networks, rather than intervening at a single or small number of targets. Combinatorial therapies could be found by searching within biomimetic pharmacological sets having the same network structure as biological systems. Evolution conducts a similar search using all controller molecules encoded in the genome, in order to find the optimal subsets to be expressed in a particular cell type. The many-to-many approach would require, rather than reject, nonspecific drugs, and may be more robust to acquired resistance and biological variability due to higher $<k_{in}>$ for drug targets. This approach to pharmacology would be modular – as in biological control – and therefore more efficient than the present practice of developing a different drug for each indication, because the same sets of molecules can be used to search for therapies for different complex diseases.

There are two recent developments that make testing this approach a realistic possibility. The first is the emergence of high-throughput *in vitro* or *in vivo* search algorithms for efficiently optimizing large combinations of drugs from within candidate sets (*17-20*). These algorithms are essential to overcome the exponentially growing possibilities of the combinatorial space. It is clearly not sufficient for pharmacological sets to have an optimal network control structure, and these methods permit an efficient search for the appropriate component drugs. The second is the availability of large libraries of suitable molecular tools, the most promising being kinase inhibitors. The 518 identified protein kinases in the human genome account for 20–30% of the drug discovery programs of many companies (*21*). There are 13 kinase inhibitors already on the market, around 200 inhibitors reported to be in clinical development, and many thousands likely to be in preclinical development. It is possible to characterize the target specificity of the inhibitors using panels of kinases (*22*).

To test the feasibility of constructing kinase inhibitor sets with biomimetic properties,

given a large library from which to choose, we simulated a drug-target network. Using a sampling algorithm, we extracted a small set of drugs having an exponential distribution of targets with $<k_{in}>$ in the biological range, and found that these inhibitors covered over 95% of targets (see online supplementary material). Therefore, pharmacological sets with biomimetic properties can be built from industrial compound libraries, if given detailed target profiles for each compound.

We have shown the generality of several network metrics of biological combinatorial control. This discovery, together with our increasing understanding of the mathematical principles underlying biological control structures and their property of efficient robustness, serve as building blocks for a completely new approach to pharmacological control of biological systems. This approach utilizes naturally occurring drug promiscuity to build sets with biomimetic properties, such as many-to-many targeting, very wide concurrent coverage of the target set, and redundancy of incoming links per target. Importantly, these are quantitative properties of the network and cannot be described simply by listing features of individual drugs, such as selectivity. This systems-level approach to pharmacological intervention is fundamentally different from the current paradigm, and would mimic combinatorial strategies that are ubiquitous in Nature.

**Acknowledgments**
This work was supported by National Science Foundation grant 0829891 and NIH grant R21AG030685. The authors declare no conflicts of interest.

## Methods

**Data and software**

Predicted human microRNA-mRNA binding sites were downloaded from the TargetScan database (*23*) release 5.1 (http://www.targetscan.org). Only conserved targets of conserved miRNA families were used (made available in the file "Predicted_Targets_Info.txt"). Human transcription factor binding sites were gathered from the TRANSFAC database (*24*). The network was trimmed for binding sites that could be mapped directly to a transcription factor with an Entrez Gene identifier (reducing 615 DNA binding domains to 389 known transcription factors and 13362 DNA binding sites to 9284 binding sites). Yeast transcription factor to gene regulations were downloaded from the YeasTRACT database (*25*) (http://www.yeastract.com). Human phosphorylation binding sites were downloaded from the PhosphoPOINT database (*26*) (http://kinase.bioinformatics.tw), using only sites in Category 3 (Known Substrate) and Category 4 (Interacting Phospho-protein with Known Substrate) (*26*). Yeast phosphorylation binding sites were extracted from the Phosphorylome database (*27*) website (http://networks.gersteinlab.org/phosphorylome/). *E. coli* transcription factor binding sites were downloaded from the RegulonDB database (*28*) release 6.4 (http://regulondb.ccg.unam.mx). Parsing and formatting of the data was performed in Python, when necessary. All data analysis was performed in R. The Bioconductor suite in R was used to perform all gene annotations ("org.Hs.eg.db" package), and Gene Ontology enrichment analysis ("GOstats package").

Numerical simulations of the mathematical model were performed in Matlab. All R and Matlab code is made available at http://paternostrolab.org/

**Degree distribution analysis**

The discrete analog to the continuous exponential distribution is the geometric distribution

$$P(X = k) = p(1-p)^{k-1}, k \in \{1, 2, ...\}$$

which has expected value

$$E(X) = \langle k \rangle = \frac{1}{p}.$$

Therefore, for a distribution with known expected value $\langle k \rangle = \mu$, $p = \frac{1}{\mu}$.

Unlike histogram approaches, the cumulative distribution function (cdf) avoids binning effects and displays every data point. In Figure 2A, empirical cumulative distribution functions for each network had their x-axis normalized by $\langle k \rangle$ and were plotted next to the cdf of the geometric distribution

$$P(X \geq k) = (1 - \frac{1}{\mu})^k,$$

with the x range normalized by μ. Similar curves were produced by different μ > 1, converging to the curve in Figure 2A for μ >> 1.

Figures 2B and 2C show binned histograms of the degree distribution data, with expected histograms of binomial distribution expected by the Erdös-Rényi random graph model. In graph theory (*29*), this model links any two nodes according to a probability *p*. Similarly, we can consider bipartite random networks of controllers and targets with the same number of control nodes *M* and target nodes *N* as each biological network, and with the probability *p* of a link between any control and any target node equal to the measured link density *D*. Random bipartite graphs have incoming and outgoing links according to the binomial distribution, using *D* as the probability parameter. Since the networks are large, the Poisson distribution

$$P(X = k) = \frac{\lambda^k e^{-\lambda}}{k!}$$

was used as an approximation to the binomial, with λ=<k>, with <k>=MD for targets and <k>=ND for controllers. The dashed curves in Figure 2B and 2C are histograms of the expected Poisson distribution of links for the M, N, and D of each network, using the same bins as the data.

**Mathematical model of a bipartite information processing network**
We neglect the feedback from targets to controllers. At the molecular level, the details of biological interactions and signal propagation are complex and idiosyncratic; therefore we used an abstract model of signaling similar to Boolean networks. In this model, control signals are represented by control node values of either 1 or 0. Links are not weighted, passing input values to the output node unaltered. Control signals reaching a target are then computed by one of three rules, and the target's output is a binary value indicating its active/inactive state. The "OR" rule designates that an output node is active if any of its connected input nodes is active. The "AND" rule requires all inputs to be active in order to activate the output node. The "MAJORITY" rule counts the number of incoming links, and activates the output node if more than half of the inputs are active, otherwise the output remains inactive. Real neurons and molecular systems often perform similar computations, and a broad spectrum of cellular reactions can be coarsely approximated by this model (*30-32*).

For a given number of controllers *M* and targets *N*, $M \leq N$, links are randomly added between controllers and targets with a probability *D*, defined as the network density, or the total links divided by the number of possible links *M*N*. Density can also be calculated from the relationship $D = \frac{\langle k_{in} \rangle}{M} = \frac{\langle k_{out} \rangle}{N}$, where <$k_{in}$> and <$k_{out}$> are the average incoming links $k_{in}$ to the *N* targets and average outgoing links $k_{out}$ from the *M* controllers, respectively.

*Robustness*

The robustness to link deletion is defined as follows: given a random bipartite network defined above, and a random binary input sequence to the controller nodes, what is the fraction of output nodes that change in response to the deletion of *γ* links? This is equivalent to asking, what is the probability that a single output node changes in response to the deleted link?

Consider a single node having $k_{in}$ incoming links and an output according to the OR rule. Define $P_F$ as the probability that a target node is in a "fragile" condition, meaning that deletion of one specific incoming link for that node will change the output. Then, the probability $F_\gamma$ that an output node changes in response (is fragile) to *γ* deleted links is

$$F_\gamma = \frac{\gamma}{L} P_F = \frac{\gamma}{\langle k_{in} \rangle N} \langle P_F(k_{in}) \rangle,$$

taking into account that the expected value of $P_F$ in a function of $k_{in}$, plus the fact that $L = \langle k_{in} \rangle N$ is the total number of links, and that *γ/L* is the probability of hitting the "fragile" link. Deleting a link to an inactive node will never change the output, so the only fragile state is to have $(k-1)$ inactive, or "0", inputs, and a single active, or "1" input, out of all $2^k$ possible binary sequences of inputs. Therefore,

$$P_F(k_{in}) = \binom{k_{in}}{1} \frac{1}{2^{k_{in}}} = \frac{k_{in}}{2^{k_{in}}},$$

and

$$R_L = 1 - F_\gamma = 1 - \frac{\mu}{\langle k_{in} \rangle} \left\langle \frac{k_{in}}{2^{k_{in}}} \right\rangle,$$

with *μ=γ/N* defined as the mutation rate of the network. Expected values for functions of $k_{in}$ are calculated below.

*Number of output states*

We define output states $\Omega(M, k_{in})$ as the total number of unique binary output sequences that our bipartite network can achieve. This quantity has a maximum of $2^M$ for a one-to-one network (see Figure 1) and a minimum of 2 for a completely connected network. We can estimate $\Omega(M,k_{in})$ for large networks by considering the output entropy for a single output node with $k_{in}$ incoming links. The number of output states is related to the single node informational entropy $S$ by

$$\Omega \approx 2^{M*S(k_{in})}, M \gg 1$$

The entropy is

$$S(k_{in}) = -q_{k_{in}}(0)\log_2 q_{k_{in}}(0) - q_{k_{in}}(1)\log_2 q_{k_{in}}(1),$$

where $q_k$ is the probability of occurrence of each output state. For the "OR" rule, only when all inputs are zero is the output also inactive, therefore

$$q_{k_{in}}(0) = \frac{1}{2^{k_{in}}}, \quad q_{k_{in}}(1) = 1 - \frac{1}{2^{k_{in}}}.$$

Inserting values for $q_k$,

$$S(k_{in}) = \frac{k_{in}}{2^{k_{in}}} - (1 - \frac{1}{2^{k_{in}}})\log_2(1 - \frac{1}{2^{k_{in}}}).$$

Using

$$\ln(1-x) \approx -\sum_{\eta=1}^{\infty}\frac{x^\eta}{\eta},$$

We obtain

$$S(k_{in}) = \frac{k_{in}}{2^{k_{in}}} + \frac{1}{\ln 2}\left(1 - \frac{1}{2^{k_{in}}}\right)\sum_{\eta=1}^{\infty}\frac{1}{\eta 2^{\eta k_{in}}},$$

and

$$\langle S(k_{in})\rangle = \left\langle\frac{k_{in}}{2^{k_{in}}}\right\rangle + \frac{1}{\ln 2}\left(1 - \left\langle\frac{1}{2^{k_{in}}}\right\rangle\right)\sum_{\eta=1}^{\infty}\left\langle\frac{1}{\eta 2^{\eta k_{in}}}\right\rangle$$

<$S(k_{in})$> can be approximated using the expected values of functions of $k_{in}$ calculated below. Plots of $\Omega(M,k_{in}) = 2^{M*<S(kin)>}$ in Figure 3 use approximations of <$S(k_{in})$> for up to $\eta$ = 3.

***Expected values***

Expected values of functions of $k_{in}$ are calculated by

$$\langle f(k_{in}) \rangle = \sum_{k=1}^{M} f(k_{in}) p(k_{in}),$$

where $p(k_{in})$ is the degree distribution of incoming links. If the links are randomly distributed with $k_{in} \in \{1, 2, \ldots M\}$, as in the network model described above, then $p(k_{in})$ is the binomial distribution. Assuming a large network, however, $p(k_{in})$ is approximated as a Poisson distribution using $k = k_{in} - 1$, and $\langle k_{in} \rangle - 1$ for the probability parameter (commonly denoted as $\lambda$), as follows

$$p(k_{in}) = \frac{(\langle k_{in} \rangle - 1)^{k_{in}-1} e^{-(\langle k_{in} \rangle - 1)}}{(k_{in} - 1)!}.$$

Using symbolic tools in Mathematica,

$$\left\langle \frac{k_{in}}{2^{k_{in}}} \right\rangle = \frac{1}{4} e^{\frac{1-\langle k_{in} \rangle}{2}} (1 + \langle k_{in} \rangle),$$

and

$$\left\langle \frac{1}{2^{\eta k_{in}}} \right\rangle = 2^{-\eta} e^{2^{-\eta}(\langle k_{in} \rangle - 1) - \langle k_{in} \rangle + 1},$$

for an integer $\eta$.

These expected values can also be calculated for an exponential (geometric) distribution, substituting $p(k_{in}) = \frac{1}{\langle k_{in} \rangle} \left(1 - \frac{1}{\langle k_{in} \rangle}\right)^{k_{in}-1}$ for the Poisson distribution. Simplifying in Mathematica, we have

$$\left\langle \frac{k}{2^{k_{in}}} \right\rangle = \frac{2 \langle k_{in} \rangle}{(1 + \langle k_{in} \rangle)^2}$$

$$\left\langle \frac{1}{2^{\eta k_{in}}} \right\rangle = \frac{1}{1 + (2^{\eta} - 1)\langle k_{in} \rangle}$$

The resulting curves are similar for Poisson and exponential link distributions (see supplementary online material), and do not affect the conclusions of the paper.

# Supplementary Online Material

**Table of Contents**

## S1. Analysis of overlap measures

We devised two measures to quantify the amount of overlap among target sets, reflecting the extent to which a set of target nodes is combinatorially regulated. "Shared Targets per Controller" (STC) is defined as the average percentage of a controller's targets that have more than one incoming link, and "Pairwise Overlap" (PO) is defined as the average percent of targets shared between any given pair of controllers (see Figure S1).

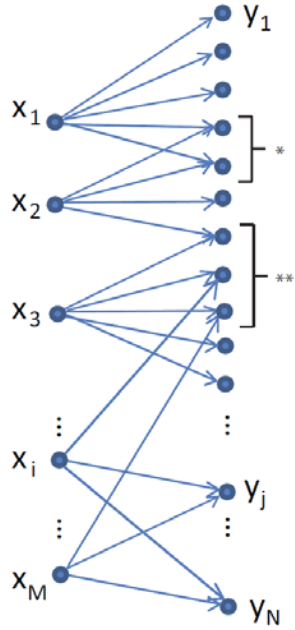

**Figure S1: Illustration of the two overlap terms** used in Table 1: *Pairwise overlap* of targets. In this example, pairwise overlap for $x_1$ with respect to $x_2$ = 2/5 (40%), ***Shared targets per controller*. In this example, the percent of shared targets for $x_3$ = 3/5 (60%)

*S1.1 Overlap measures in biological networks*

All networks had high STC, meaning that any control node shares the vast majority of its targets with at least one other controller. The mean PO between any two controllers was an average of 5.7% ± 2.6% over all networks (Table 1 in the text). Using the random graph framework, we derived the following equations for the expected values of STC and PO in bipartite random networks as a function of *M*, *N*, and *D* (derivations in the next section):

$$E[STC] = [1 - (1 - D)^N]*[1 - (1 - D)^{M-1}],$$
$$E[PO] = [1 - (1 - D)^N]*D.$$

For each biological network, we then compared the calculated overlap parameters to the values expected for random graphs of equal size. We also generated shuffled versions of the networks by swapping links while retaining the original link-per-node distributions (Table S1). We found that average values for Shared Targets per Controller were significantly lower in biological systems than in their random counterparts. Pairwise Overlap was significantly higher in the biological networks except for the *E. coli* transcription factors. Both metrics were mostly explained by the link distributions, as evidenced by the similarities between the biological and the shuffled networks.

**Table S1: Comparison of overlap parameters in biological networks to random.** Shuffled networks have equal numbers of nodes and links, as well as equivalent link distributions, as the biological network. Random networks have equal numbers of nodes and links, with links placed randomly. Shared Targets per Controller (STC) and Pairwise Overlap (PO) measurements are presented as mean over all controllers or pairs of controllers, respectively. Mean and standard deviation of overlap parameters in the shuffled and random networks varied less than 2% over 5 simulations.

**Shared Targets per Controller (STC)**

| Network | Biological | | Shuffled (equal link dist.) | | Random (equal # nodes, links) | |
|---|---|---|---|---|---|---|
| *Average over controllers* | Mean | Std dev | Mean | Std dev | Mean | Std dev |
| Human TF | 97.8% | 2.4% | 97.7% | 1.4% | 100.0% | 0.2% |
| Human miRNA | 95.3% | 3.0% | 98.7% | 0.9% | 99.5% | 0.4% |
| Human Kinase | 73.5% | 34.3% | 79.7% | 27.1% | 90.3% | 10.7% |
| Yeast TF | 98.1% | 4.3% | 98.7% | 1.4% | 99.9% | 0.2% |
| Yeast Kinase | 86.0% | 14.3% | 87.5% | 9.9% | 94.8% | 3.2% |
| E. coli TF | 73.5% | 36.1% | 83.4% | 17.3% | 88.9% | 7.5% |

**Pairwise shared targets (PO)**

| Network | Biological | | Shuffled (equal link dist.) | | Random (equal # nodes, links) | |
|---|---|---|---|---|---|---|
| *Average over controllers* | Mean | Std dev | Mean | Std dev | Mean | Std dev |
| Human TF | 4.5% | 1.0% | 4.5% | 0.4% | 1.9% | 0.2% |
| Human miRNA | 7.1% | 1.1% | 7.3% | 0.9% | 3.4% | 0.1% |
| Human Kinase | 1.5% | 1.3% | 1.6% | 1.4% | 0.9% | 0.2% |
| Yeast TF | 6.3% | 1.8% | 6.5% | 1.2% | 3.5% | 0.1% |
| Yeast Kinase | 8.5% | 3.7% | 9.0% | 3.2% | 3.4% | 0.3% |
| E. coli TF | 1.1% | 0.9% | 1.5% | 0.7% | 1.3% | 0.2% |

*S1.2 Derivation of overlap measures in random networks*

Let us consider an ensemble of bipartite directed network characterized by $M$ controllers and $N$ targets. Many properties of the ensemble can be defined in terms of the distribution for the number of links out of a controller node, $p(N,k_{out})$, and the distribution for the number of links into a target node, $q(M,k_{in})$. We will focus on the Erdös-Rényi random graph model (*1*), for which these two distributions read

$$p(N,k_{out}) = \binom{N}{k_{out}} D^{k_{out}} (1-D)^{N-k_{out}},$$

$$q(M, k_{in}) = \binom{M}{k_{in}} D^{k_{in}} (1-D)^{M-k_{in}},$$

where $D$ is the probability to have a link.

The Shared Targets per Controller (STC) is a quantity that characterizes how likely a target reached by a controller is shared with other controllers. This can be calculated by fixing one particular controller out of the $M$ available and defining the quantity

$$z = \sum_{k_{in}=1}^{M-1} q(M-1, k_{in}).$$

The quantity $z$ is the total probability that at least one of the other $M - 1$ controllers is co-controlling a target reached by the initially fixed controller. If the initial controller has $k_{out}$ links out, then the Shared Targets per Controllers is

$$STC(k_{out}) = \frac{1}{k_{out}} \sum_{m=1}^{k_{out}} m \binom{k_{out}}{m} z^m (1-z)^{k_{out}-m},$$

where $1/k_o$ is a normalization factor, and the binomial coefficient takes into account in how many ways $k_o$ targets can be co-controlled by additional $m$ controllers. The final expression for the STC can be written as

$$STC = \sum_{k_{out}}^{N} p(N, k_{out}) STC(k_{out}),$$

which in the random model leads to the simple expression

$$STC = \left[1 - (1-D)^N\right]\left[1 - (1-D)^{M-1}\right],$$

since $STC(k_{out}) = z$ does not depend on $k_{out}$ in this case.

The Pairwise Overlap (PO) is a quantity that characterizes the probability that two controllers overlap by acting on the same target. This quantity can be calculated in a similar way as the STC by fixing a pair of controller nodes. Then we can define the quantity

$$h = \sum_{k_{in}=1}^{M-1} q(M-1, k_{in}) \frac{k_{in}}{M-1},$$

which gives the total probability that one node at the end of the link from the first controller is also connected to the second controller in the pair. If the initial controller has $k_{out}$ links out, then the Pairwise Overlap can be explicitly expressed in terms of $h$ as

$$PO(k_{out}) = \frac{1}{k_{out}} \sum_{m=1}^{k_{out}} m \binom{k_{out}}{m} h^m (1-h)^{k_{out}-m},$$

and the total PO is

$$PO = \sum_{k_{out}=1}^{N} p(N, k_{out}) PO(k_{out}).$$

In the random graph case we have $PO(k_{out}) = h = D$, which leads to the final result

$$PO = \left[1 - (1-D)^N\right] D.$$

## S2. Further analysis of link distributions

*S2.1 Fitting incoming/outgoing links to exponential and scale-free distributions*

Cumulative distributions of links per node were compared to each other in Figure 2A of the main text, while histograms of incoming and outgoing links in the biological control networks were compared against exponential and binomial distributions in Figure 2B and 2C. Here we fit the empirical cumulative distribution function for each individual dataset against both exponential and scale-free distributions, and measure the goodness of fit. Figures S2 through S5 use a log-transformed y-axis to plot the cumulative distribution function (cdf) and log-transformed cdf, in order to measure fit to an exponential and scale-free distribution, respectively.

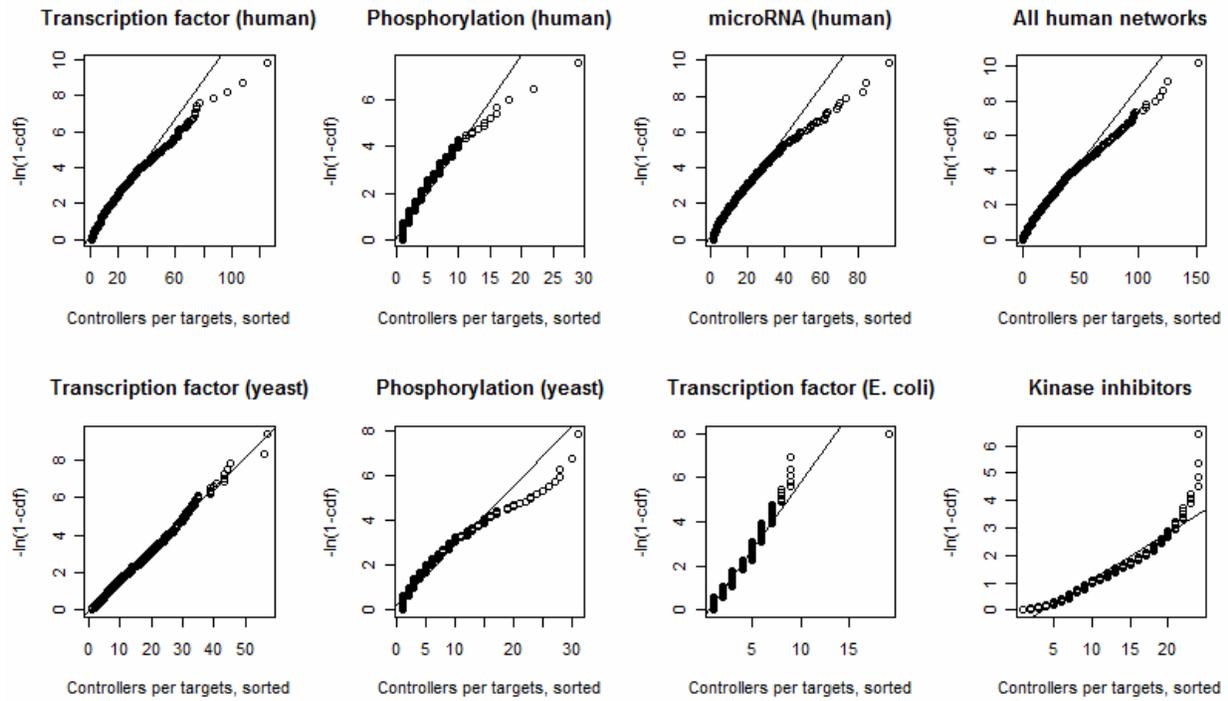

**Figure S2**

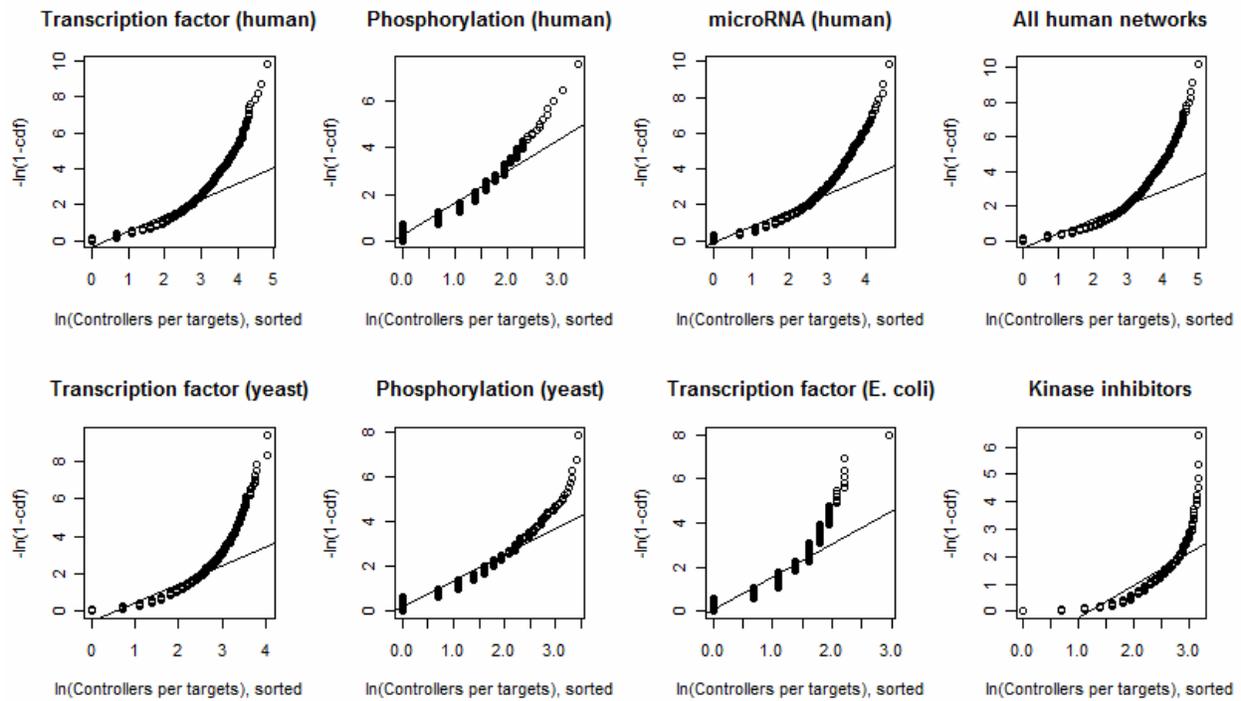

**Figure S3**

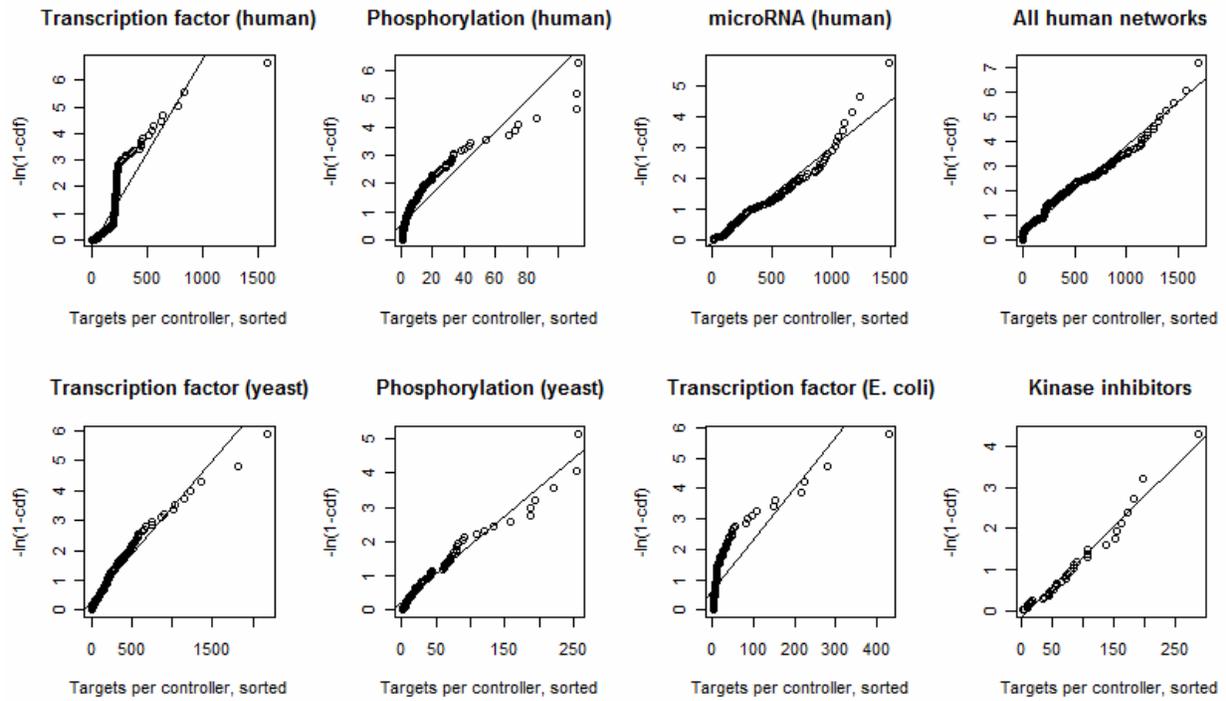

**Figure S4**

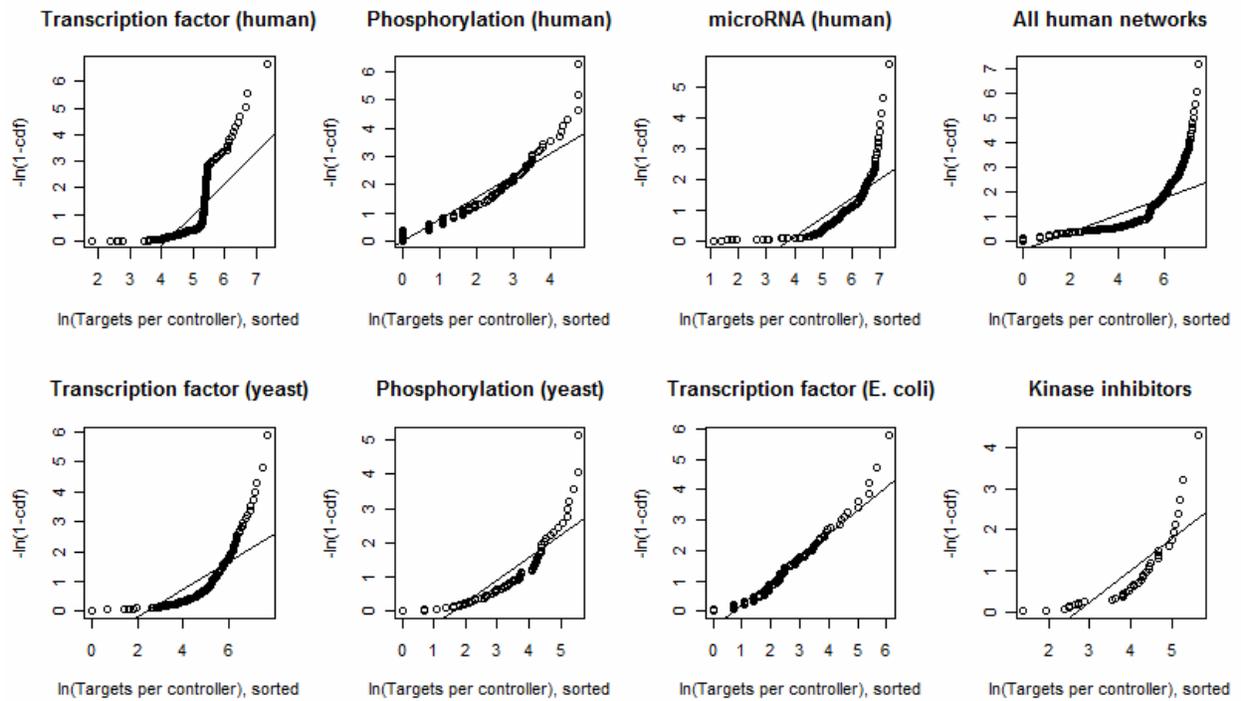

**Figure S5**

**Figure S2: Fitting controllers per target (incoming links) to an exponential distribution.** The E. coli and yeast transcription factor networks fit tightly with this distribution, while all human networks and the yeast phosphorylation network seem to have a fat-tail or scale-free component

**Figure S3: Fitting controllers per target (incoming links) to a scale-free distribution.** The human and yeast phosphorylation networks fit more tightly with this distribution.

**Figure S4: Fitting targets per controller (outgoing links) to an exponential distribution.** All but the *E. coli* transcription factor network have at least some exponential component.

**Figure S5: Fitting targets per controller (outgoing links) to a scale-free distribution.** The *E. coli* transcription factor network is better modeled by a scale-free distribution, and the human kinase network may also have a scale-free component.

The microRNA and all three transcription factor networks seem to better fit an exponential distribution, while the phosphorylation networks in both human and yeast may have some scale-free component. The yeast and E. coli transcription factor networks are well-modeled by exponential distributions. Table S2 compares all fits using $R^2$ values. For both types of links, the human kinase network is a better fit to a scale-free distribution. With the exception of the human transcription factor network, all networks and link types fit at least one distribution with an $R^2$ of > 0.9.

**Table S2: Evaluation of fitting models.** Higher R-squared values for each network are in bold.

Controllers per target (incoming)

| **Exponential fit** | | | **Scale-free fit** | | |
|---|---|---|---|---|---|
| *Network* | *slope* | *R squared* | *Network* | *slope* | *R squared* |
| Human miRNA | 13.2 | **0.964** | Human miRNA | 0.92 | 0.881 |
| Human TF | 10.1 | **0.967** | Human TF | 0.88 | 0.867 |
| E. coli TF | 9.2 | **0.953** | E. coli TF | 1.49 | 0.91 |
| Yeast TF | 10.4 | **0.997** | Yeast TF | 1.01 | 0.753 |
| Human Kinase | 3.9 | 0.916 | Human Kinase | 1.37 | **0.939** |
| Yeast Kinase | 3.6 | 0.92 | Yeast Kinase | 1.18 | **0.958** |

Targets per control (outgoing)

| **Exponential fit** | | | **Scale-free fit** | | |
|---|---|---|---|---|---|
| *Network* | *slope* | *R squared* | *Network* | *slope* | *R squared* |
| Human miRNA | 0.0047 | **0.964** | Human miRNA | 0.62 | 0.579 |
| Human TF | 0.028 | **0.74** | Human TF | 1.18 | 0.567 |
| E. coli TF | 0.028 | 0.697 | E. coli TF | 0.77 | **0.928** |
| Yeast TF | 0.0059 | **0.948** | Yeast TF | 0.47 | 0.6055 |
| Human Kinase | 0.15 | 0.837 | Human Kinase | 0.78 | **0.911** |
| Yeast Kinase | 0.015 | **0.9646** | Yeast Kinase | 0.65 | 0.758 |

*S2.2. Exponential distributions cannot be obtained by sampling a scale-free network*

To detect the possibility that the exponential distribution was an artifact of sampling from a scale-free network, we generated a scale-free network in-silico using the Barabasi-Albert (B-A) growth and preferential attachment model (*2*), and then randomly assigned nodes to control and target layers, keeping all links between the two layers. A bipartite network was also generated de-novo by constraining the B-A model to operate with two node classes and unidirectional links. In both cases, the resulting bipartite networks followed a scale-free distribution rather than an exponential.

*S2.3 Effect of link distribution on the analytical network model*

As shown in the methods, analytical expressions for robustness and output states (as a function of the entropy of the output node) can be computed for different distributions of links. Since the biological networks are shown to be exponentially distributed, it is important to examine the effect of the distribution on the relationship to <$k_{in}$>. In Figure S6, we show that the curves are similar for the exponential and Poisson distributions, and do not alter the main conclusions of the paper.

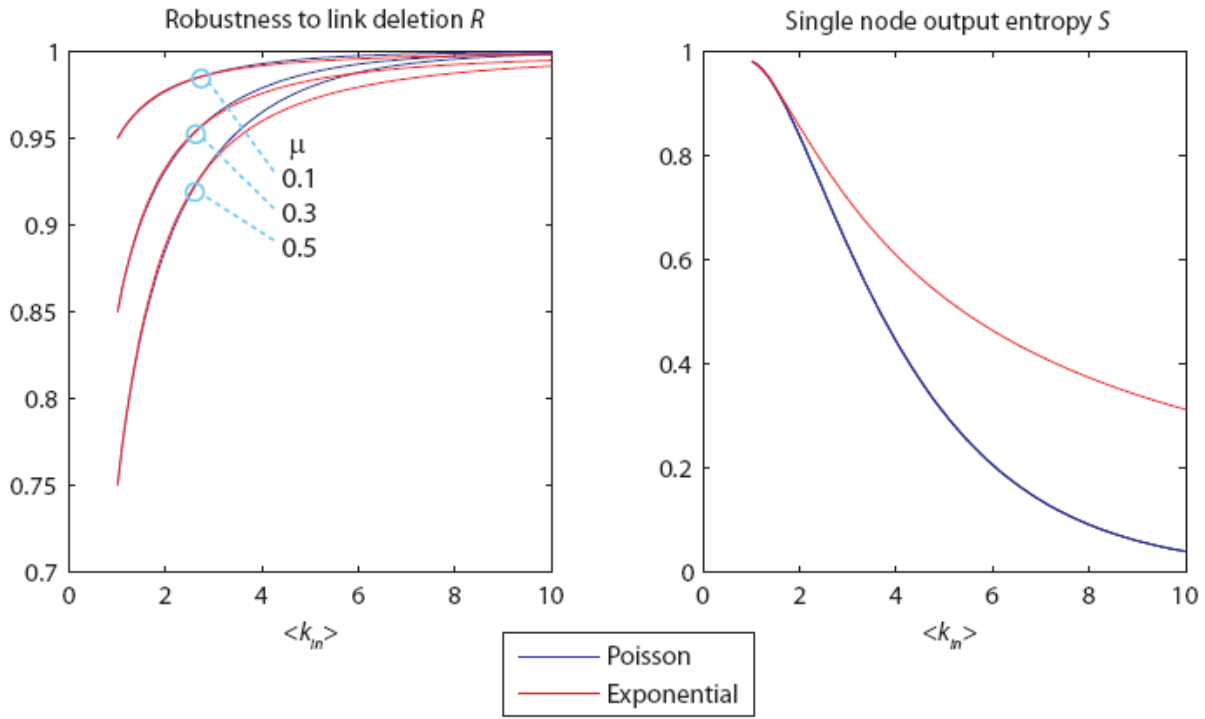

**Figure S6: Comparison of the analytical model of Figure 3 for the two different link distributions**

## S3. Enrichment of gene categories in network targets

Since target nodes have a broad distribution of incoming control links, we used the three human networks to explore whether certain categories of nodes may be more highly targeted than others. Controller nodes appeared in the target sets more than expected by random (Section S3.1). Highly targeted genes in all networks shared many significantly enriched Gene Ontology (*3*) (GO) terms involved in transcription, regulation, and development (Section S3.2). Conversely, sparsely targeted genes tended to be enriched in GO terms involving biological "effector" processes, such as metabolism, transport, and the response to stimulus. Additionally, human genes regulated by all three types of controller molecule were almost always themselves involved in regulation (Section S3.3). Together these data suggest that cells use different control network topologies depending on the type of target genes. Control nodes themselves are under the heaviest combinatorial control, and by more different types of controller, while downstream effector genes are regulated by fewer controllers.

*S3.1 Enrichment of controller nodes in target sets by network*

We examined the enrichment of controller nodes from each human network within all target sets. Kinases and transcription factors can be targets for regulation by all three controller networks, whereas microRNAs are not directly regulated by other microRNAs and cannot be phosphorylated by kinases (NA in Table S1). MicroRNAs are known to be regulated by transcription factors, but to our knowledge large-scale binding data are not available. Transcription factors in TRANSFAC were mapped to Entrez gene numbers where possible by both automatic and manual methods, resulting in a trimmed list of 197 transcription factors. All kinases were mapped to Entrez gene identifiers in the original PhosphoPOINT database. Table S2 shows the representation of kinases and transcription factors as targets in the three networks.

Enrichment of transcription factors and kinases (control nodes from their respective network databases) in target sets of each of the networks were found to be significant by hypergeometric tests in a gene universe of 20,500, resulting in p-values below the computational lower limit in each case. The analysis was repeated against the full set of transcription factors and kinases in the human genome, as identified by GO annotation (both GO:0003677 – "DNA binding" and GO:0003700 – "transcription factor activity" for transcription factors; and GO: 0004672 – "protein kinase activity" for kinases), and these gene sets were similarly found to be significantly enriched in all target sets in all cases.

**Table S3: Presence of controller nodes of each type in the target sets of human networks**. Transcription factors and kinases were significantly enriched in the target sets of all three networks. For example, of the 389 transcription factors from the human TF network, 147 were found in the target set of the miRNA network.

| Network | Number of controllers in target set (total controller pool) | | | |
|---|---|---|---|---|
| | *miRNA (153)* | *TF (389)* | *Kinase (264)* | *All targets* |
| *miRNA* | NA | 147 | 206 | 9448 |
| *TF* | no data | 151 | 201 | 9284 |
| *Kinase* | NA | 55 | 167 | 988 |

*S3.2 Unbiased GO enrichment of target nodes separated by $k_{in}$*

For each network, we divided target nodes into two subsets based on the number of linked controllers: "highly targeted" genes with greater than 5 times the mean incoming links of the network, and "sparsely targeted" genes with less than half the mean incoming links of the network. Each subset was then subjected to unbiased Gene Ontology (GO) enrichment analysis to find over-represented gene categories. The top 10 GO terms that were enriched in the highly-targeted subset are shown in Table S3. Over-represented GO terms for the sparsely-targeted subset of genes are presented in Table S4. Here, non-coding RNA metabolism appeared in the top GO terms for both the microRNA and transcription factor networks. Oxidation reduction was the only other shared term, while the phosphorylation network did not have any significantly over-represented terms in its low-degree target proteins (p-value threshold = 0.001). As discussed in the main text, GO terms involved in "regulation" were more prevalent in the highly-targeted set, while genes that can be thought of as "effectors" (e.g. metabolic genes, transporters, and response genes) were enriched in the sparsely-targeted set.

**Table S4: Top 10 over-represented GO Biological Process terms for highly targeted genes** in three human networks (genes with incoming links greater than 5 times the network average). Bold denotes appearance in more than one network (even if not shown in the top 10). Size is the number of target genes in both subsets that are associated with the GO term. ExpCount is expected number of appearances of the term and Count is the actual number of appearances

*Highly targeted: > 5*mean(controls per target)*

| GOBPID | Pvalue | ExpCount | Count | Size | Term |
|---|---|---|---|---|---|
| *TF* | | | | | |
| GO:0080090 | 1.0E-59 | 227.5 | 441 | 1655 | **regulation of primary metabolic process** |
| GO:0060255 | 3.4E-58 | 230.7 | 442 | 1678 | **regulation of macromolecule metabolic process** |
| GO:0032774 | 1.7E-50 | 127.8 | 290 | 930 | **RNA biosynthetic process** |
| GO:0034961 | 8.6E-49 | 233.3 | 426 | 1697 | **cellular biopolymer biosynthetic process** |
| GO:0065007 | 2.1E-47 | 524.4 | 732 | 3815 | biological regulation |
| GO:0009059 | 4.0E-43 | 249.5 | 433 | 1815 | **macromolecule biosynthetic process** |
| GO:0009889 | 3.5E-42 | 113.6 | 253 | 907 | **regulation of biosynthetic process** |
| GO:0010467 | 9.9E-42 | 262.4 | 445 | 1909 | **gene expression** |
| GO:0051171 | 1.7E-41 | 110.3 | 247 | 882 | **regulation of nitrogen compound metabolic process** |
| GO:0031323 | 3.3E-41 | 128.9 | 272 | 1037 | **regulation of cellular metabolic process** |
| | | | | | |
| *microRNA* | | | | | |
| GO:0060255 | 1.9E-16 | 267.3 | 377 | 1896 | **regulation of macromolecule metabolic process** |
| GO:0080090 | 3.0E-16 | 268.8 | 378 | 1907 | **regulation of primary metabolic process** |
| GO:0009889 | 3.2E-15 | 239.4 | 341 | 1698 | **regulation of biosynthetic process** |
| GO:0031323 | 4.9E-14 | 280.5 | 382 | 1990 | **regulation of cellular metabolic process** |
| GO:0051171 | 1.4E-13 | 230.9 | 325 | 1638 | **regulation of nitrogen compound metabolic process** |
| GO:0044260 | 4.6E-12 | 460.7 | 562 | 3268 | **cellular macromolecule metabolic process** |
| GO:0050789 | 8.7E-12 | 534.4 | 634 | 3791 | regulation of biological process |
| GO:0034961 | 2.6E-11 | 254.2 | 341 | 1803 | **cellular biopolymer biosynthetic process** |
| GO:0009059 | 1.5E-08 | 213.6 | 282 | 1556 | **macromolecule biosynthetic process** |
| GO:0043283 | 6.4E-08 | 328.2 | 399 | 2422 | **biopolymer metabolic process** |
| | | | | | |
| *Kinase* | | | | | |
| GO:0034961 | 4.9E-04 | 3.3 | 9 | 281 | **cellular biopolymer biosynthetic process** |
| GO:0009059 | 6.1E-04 | 3.3 | 9 | 288 | **macromolecule biosynthetic process** |
| GO:0044419 | 8.2E-04 | 0.9 | 5 | 74 | interspecies interaction between organisms |
| GO:0006355 | 9.7E-04 | 2.0 | 7 | 171 | **regulation of transcription, DNA-dependent** |

**Table S5: Top 10 over-represented GO Biological Process terms for low-degree genes** in three human networks (genes with incoming links less than 2 times the network average). Bold denotes appearance in more than one network. Size is the number of target genes in both subsets that are associated with the GO term.

**Low targeted: 0.5*mean(controls per target)**

| GOBPID | Pvalue | ExpCount | Count | Size | Term |
|---|---|---|---|---|---|
| *TF* | | | | | |
| GO:0034660 | 1.1E-05 | 40.51129 | 62 | 101 | **ncRNA metabolic process** |
| GO:0022900 | 1.2E-05 | 27.27493 | 45 | 68 | electron transport chain |
| GO:0055114 | 2.1E-05 | 91.28796 | 122 | 229 | **oxidation reduction** |
| GO:0044271 | 6.4E-05 | 33.29146 | 51 | 83 | nitrogen compound biosynthetic process |
| GO:0042773 | 7.8E-05 | 13.63747 | 25 | 34 | ATP synthesis coupled electron transport |
| GO:0042180 | 8.4E-05 | 111.1052 | 142 | 277 | cellular ketone metabolic process |
| GO:0006120 | 9.7E-05 | 11.63196 | 22 | 29 | mitochondrial electron transport, NADH to ubiquinone |
| GO:0006968 | 9.7E-05 | 11.63196 | 22 | 29 | cellular defense response |
| GO:0006732 | 1.5E-04 | 27.67603 | 43 | 69 | coenzyme metabolic process |
| GO:0022613 | 1.6E-04 | 31.68705 | 48 | 79 | ribonucleoprotein complex biogenesis |
| | | | | | |
| *miRNA* | | | | | |
| GO:0055114 | 2.5E-10 | 97.46176 | 145 | 240 | **oxidation reduction** |
| GO:0034660 | 6.4E-10 | 27.20807 | 52 | 67 | **ncRNA metabolic process** |
| GO:0007600 | 2.0E-06 | 70.65978 | 101 | 174 | sensory perception |
| GO:0006955 | 5.7E-06 | 85.33713 | 117 | 211 | immune response |
| GO:0022613 | 1.3E-05 | 25.17762 | 42 | 62 | ribonucleoprotein complex biogenesis |
| GO:0006396 | 4.4E-05 | 109.2384 | 141 | 269 | RNA processing |
| GO:0006364 | 5.8E-05 | 8.121813 | 17 | 20 | rRNA processing |
| GO:0007601 | 6.3E-05 | 42.23343 | 62 | 104 | visual perception |
| GO:0006091 | 1.2E-04 | 60.50751 | 83 | 149 | generation of precursor metabolites and energy |
| GO:0045087 | 0.0007 | 21.11671 | 33 | 52 | innate immune response |

*Kinase*
none

### S3.3 Enrichment of target nodes by number of controller types

Next, we separated genes by the number of different *types* of control molecule targeting a given gene. Figure S6 is a Venn diagram of all human genes with GO annotations, separated into groups targeted by 0, 1, 2 or all 3 of the cellular control networks. Genes not targeted by any network were enriched in GO annotations involved in neurological, immune, and G-protein signaling systems, which might be more indicative of our incomplete knowledge of these networks. Genes targeted by all three types of controller molecule were often involved in regulation themselves, much like the highly targeted genes in Table S4. Genes targeted by only one of the miRNA, transcription factor, or kinase networks have unique properties that are often related to the type of controller, for example genes targeted only by miRNAs are more likely to be involved in RNA splicing, and kinase targets are enriched in genes annotated as "intracellular signaling cascade."

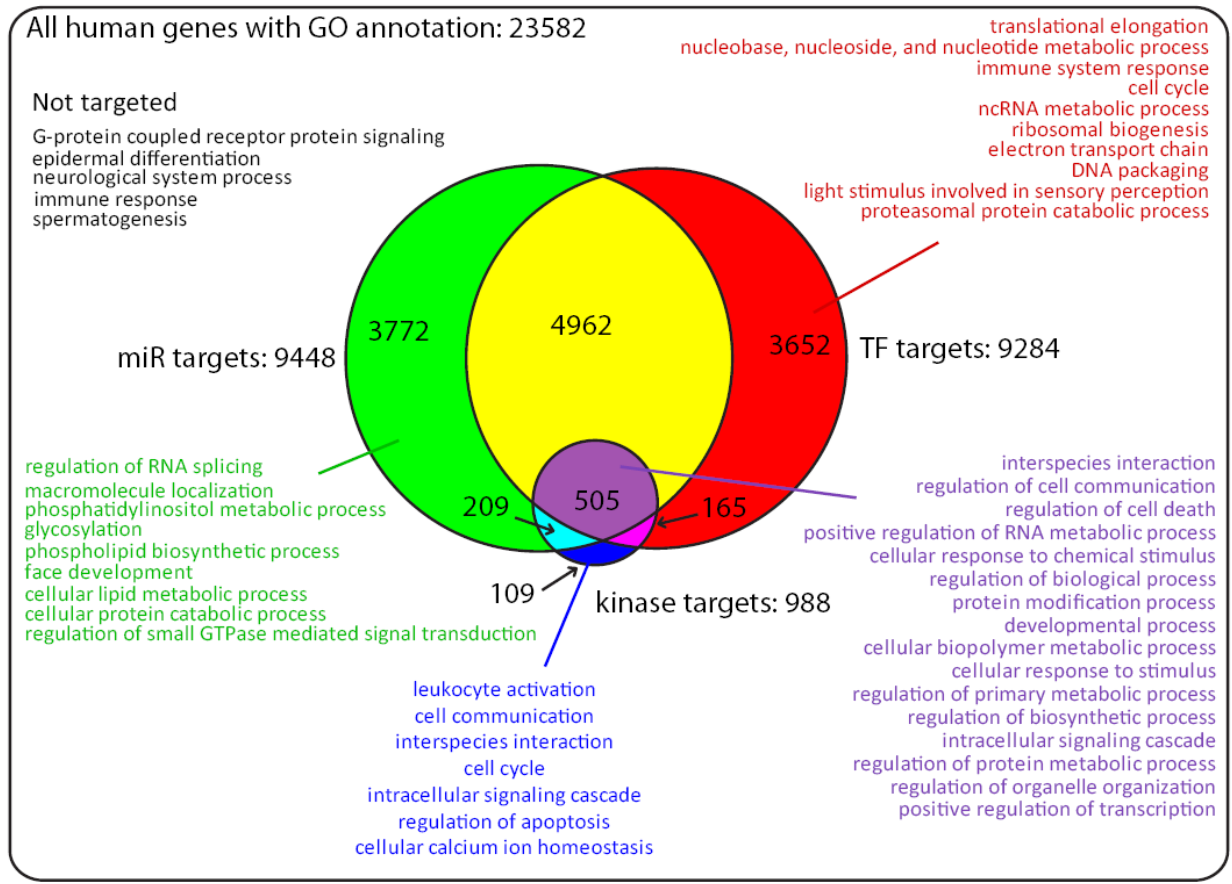

**Figure S7: Venn diagram of human gene targets, by types of controller molecule.** Selected top GO annotations (p-value < 0.001) for each slice of the Venn diagram are listed.

## S4. Additional examples of many-to-many control in biology

A many-to-many combinatorial structure is not limited to the control of cells and it is found in all types of complex control in biology, the most striking example being the control of the organism by the nervous system, where connections among neurons have a many-to-many arrangement. The control of effectors by neurons has a simpler structure, as shown by motor units (*4*), where each motor neuron controls a distinct set of muscle fibers and the target sets are not overlapping, in a one-to-many fashion. The complexity of control structure might depend on the complexity of the target system.

Diseases such as cancer may also adapt by developing combinatorial strategies to counter intrinsic defense mechanisms and homeostatic reactions or extrinsic therapeutic interventions (*5*). An increasing body of evidence shows that the resistance of cancer to therapies involves molecules acting at multiple levels with many-to-many actions. This provides further support for the use of biomimetic therapeutic strategies of matching complexity.

## S5. Sampling a simulated kinase inhibitor library to obtain a biomimetic set

Drug companies often have hundreds or thousands of kinase inhibitors with known targeting profiles. We simulated a much larger library of 1500 compounds, and created target profiles that gave the same target per controller and controller per target distributions as the 38-drug network in Karaman et. al. (*6*)

The simulated library was created using the inverse sampling transform method, which requires the analytic inversion of the cumulative distributions of the theoretical distributions we want to sample (*7*). This method is used both for targets and for controllers. A link-matching procedure is then implemented to randomly match "links out" of kinase inhibitors with "links in" into kinase nodes, creating a bipartite network with the desired controllers per target and targets per controller distributions. We show in Figure S7 the targets per controller and controlled per target distribution for a simulated network obtained with this procedure. The desired distributions are those of the kinase inhibitor network in Figure 2B and 2C of the main text.

Once a sample kinase inhibitor/kinase network has been created, we have used a rejection method approach (*7*) to identify a subset of inhibitors having an exponential distribution, but a reduced average value for <$k_{out}$>, according to the biomimetic criterion. The rejection method consists in picking randomly an inhibitor node with a $k_{out} = k$, and keeping it in the set with probability $p(k) = \dfrac{1}{k_{out,BM}} \left(1 - \dfrac{1}{k_{out,BM}}\right)^{(k-1)}$, where $k_{out,BM}$ is the ideal biomimetic value. We have tested this procedure for different initial size of the KI library, and we observed that for a large enough library we obtain a very high coverage of all kinases, as expected from the many-to-many structure of biomimetic sets.

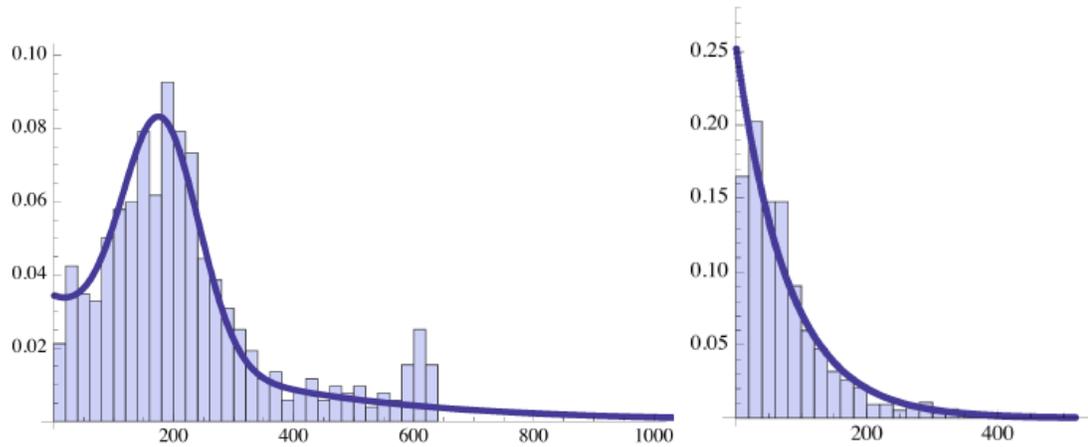

**Figure S8: Targets per controller and controllers per target for the simulation of a large kinase inhibitor library.**

## S6. Supplementary References

1. M. E. J. Newman, S. H. Strogatz, D. J. Watts, *Phys. Rev. E* **64** (2001).

2. Barabasi, Albert, *Science* **286**, 509-512 (1999).

3. M. Ashburner et al., *Nat. Genet* **25**, 25-29 (2000).

4. R. J. Monti, R. R. Roy, V. R. Edgerton, *Muscle Nerve* **24**, 848-866 (2001).

5. J. Zhou, *Multi-Drug Resistance in Cancer* (Springer-Verlag GmbH, 2009).

6. M. W. Karaman et al., *Nat. Biotechnol* **26**, 127-132 (2008).

7. W. H. Press, *Numerical recipes: the art of scientific computing* (Cambridge University Press, 2007).